\documentclass[twocolumn,aps,pra,showpacs,groupedaddress,superscriptaddress,nofootinbib]{revtex4-1}

\usepackage[pdftex,linkcolor=blue,citecolor=blue,urlcolor=blue,colorlinks]{hyperref}
\usepackage[dvipsnames]{xcolor}
\usepackage{graphicx,epsfig}
\usepackage{braket}
\usepackage{color}
\usepackage{cancel}
\usepackage{amsmath}
\usepackage{amsfonts}
\usepackage{fancyhdr}

\begin{document}
\title{Single atom edge-like states via quantum interference}
\author{G. Pelegr\'i}
\affiliation{Departament de F\'isica, Universitat Aut\`onoma de Barcelona, E-08193 Bellaterra, Spain.}
\author{J. Polo}
\affiliation{Departament de F\'isica, Universitat Aut\`onoma de Barcelona, E-08193 Bellaterra, Spain.}
\author{A. Turpin}
\affiliation{Departament de F\'isica, Universitat Aut\`onoma de Barcelona, E-08193 Bellaterra, Spain.}
\author{M. Lewenstein}
\affiliation{ICFO - Institut de Ci\`encies Fot\`oniques, The Barcelona Institute of Science and Technology, Av. C.F. Gauss 3, 08860 Castelldefels (Barcelona), Spain.}
\affiliation{ICREA - Instituci\'{o} Catalana de Recerca i Estudis Avan\c{c}ats, Lluis Companys 23, 08010 Barcelona, Spain.}
\author{J. Mompart}
\affiliation{Departament de F\'isica, Universitat Aut\`onoma de Barcelona, E-08193 Bellaterra, Spain.}
\author{V. Ahufinger}
\affiliation{Departament de F\'isica, Universitat Aut\`onoma de Barcelona, E-08193 Bellaterra, Spain.}

\begin{abstract}
We demonstrate how quantum interference may lead to the appearance of robust edge-like states of a single ultracold atom in a two-dimensional optical ribbon. We show that these states can be engineered either within the manifold of local ground states of the sites forming the ribbon, or of states carrying one unit of angular momentum. In the former case, we show that the implementation of edge-like states can be extended to other geometries, such as tilted square lattices. In the latter case, we suggest to use the winding number associated to the angular momentum as a synthetic dimension.   
\end{abstract}
\pacs{03.75.Lm, 67.85.-d}
\maketitle


\section{Introduction}
Ultracold atoms in optical lattices provide an ideal playground for studying condensed matter phenomena in a highly controlled and tunable manner \cite{llibreveronica, reviewbloch}. In recent years, the realization of artificial gauge fields \cite{jaksch_zoller, osterloh2005,mueller,gerbier_dalibard,eckert2010,struck2012,hauke2012, gediminasRMP,goldman_etal_review,shakingreview} has opened the possibility to use these systems to explore physics of strong magnetic fields \cite{chernumber, BECstrongfield, currentbosons, hofstader, harper, edgebosons, edgefermions}. In particular, edge states predicted in the context of the quantum Hall effect have been observed both with bosons \cite{edgebosons} and fermions \cite{edgefermions} in one-dimensional (1D) optical lattices extended in a synthetic dimension by taking profit of the internal atomic degrees of freedom \cite{syntheticdimensions}. The robustness of these states makes them useful for instance for quantum information purposes \cite{edgequantuminfo, edgequantuminfo2}.

Here, we propose a scheme to generate robust edge-like states (ELS) in two-dimensional (2D) arrangements of discrete sites without the need to create synthetic gauge fields. The method is based on the use of spatial dark states (SDS), which appear on tunneled-coupled three-site systems due to quantum interference \cite{spatialdark} and are the basis for spatial matter-wave passage techniques \cite{reviewsap}. We focus on single atoms or non-interacting Bose-Einstein condensates. 

SDS can be realized using states carrying orbital angular momentum (OAM) $l\in \mathbb{Z}$. In this work we focus on the use of states with $l=0$ and $l=1$. In the $l=0$ case, quantum interference effects are solely due to phase differences in the local states of the sites, allowing to create ELS in a large variety of geometrical configurations of sites. These arrangements of sites could be realized using, for instance, painting potentials \cite{painting1, painting2, painting3, painting4} or spatial light modulators \cite{slm1, slm2}. In the $l=1$ case, quantum interference is also due to complex tunneling amplitudes, whose phases are modulated by the relative orientation between sites \cite{geometricallyinduced}. The manifold of $l=1$ states, which has been object of intensive study in recent years \cite{l1,l2,l3,l4,l5,l6,l7,l8,l9,l10,l11}, offers the additional possibility of using the winding number as a synthetic dimension. This could open the door to the quantum simulation of non-trivial topologies \cite{nontrivialtopology,  synthdimphotonic}, with the advantage that complex tunneling amplitudes appear naturally.

The paper is organized as follows. First, in Sec.~\ref{sec1} we describe the physical system in which we shall implement the ELS, and we argue that the dynamics of a single atom can be studied separately for each manifold of OAM states. Then, in Sec.~\ref{sec2} we discuss the form and the properties of the ELS, both for the manifold of $l=0$ states (Sec.~\ref{sec2a}) and that of $l=1$ states (Sec.~\ref{sec2b}). Finally, in Sec.~\ref{sec3} we expose some conclusions.
\section{Physical system}
\label{sec1}
\begin{figure}[t!]
\centering
\includegraphics[width=0.8\linewidth]{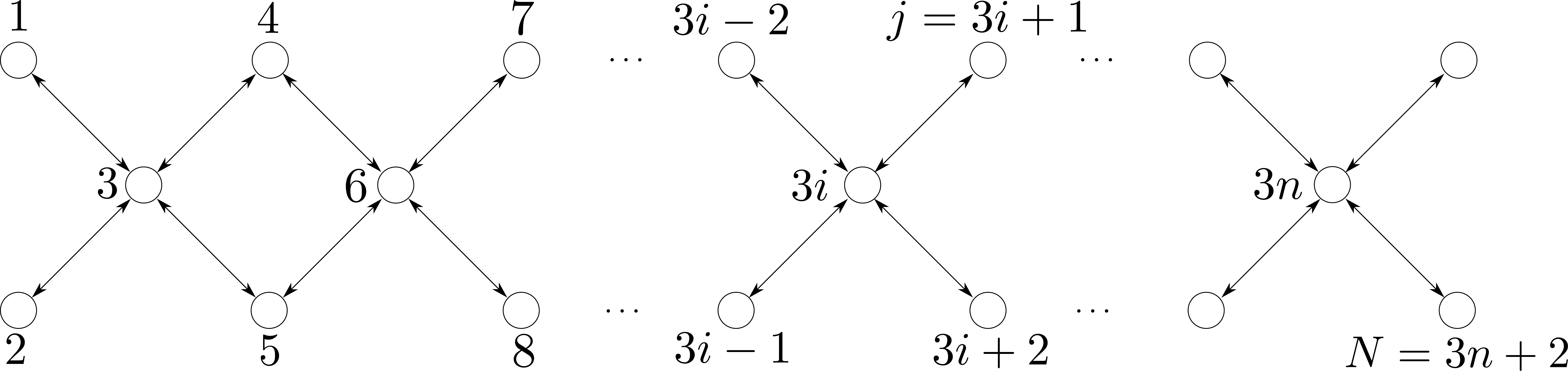}
\caption{Sketch of the considered 2D optical ribbon, with circles representing its sites. The index $i=1,...,n$ refers to individual cells, which are squares with one site laying at each of their corners plus a central site placed at an equal distance $d$ from the outer ones. The sites are labelled with the index $j=1,...,N$; with $N=3n+2$. As indicated by the double arrows, we only consider the central sites to be tunneled-coupled to their four nearest neighbour sites.}
\label{esquemaribbon}
\end{figure}

A simple system in which we can take advantage of quantum interferences to implement single atom ELS is shown in Fig.~ \ref{esquemaribbon}. It is a 2D ribbon constructed by placing side by side five-site cells consisting of a square with a site at each corner plus a central site laying at a distance $d$ from the other four. Thus, a lattice with $n$ cells has a total of $N=3n+2$ sites. We assume that each site hosts a harmonic trap of identical frequency $\omega$. If the sites are separated enough, their local eigenfunctions can be used as a basis of the total Hilbert space associated to the lattice. Moreover, the total set of eigenstates can be split into manifolds of total OAM $l$, each containing $N\cdot(l+1)$ degenerate states corresponding to the eigenstates with $z$ component of the OAM $m=-l,-l+2,...,l-2,l$ in each site. Therefore, under this assumption the dynamics of the system can be studied separately for each OAM manifold, leading to a total Hamiltonian  
\begin{equation}
\hat{H}_{\text{ribbon}}=\sum_{l=0}^{\infty}\hat{H}_l,
\end{equation}
where $\hat{H}_l$ is a few-state Hamiltonian describing the tunneling dynamics within the manifold of OAM $l$.
\section{Edge-like states}
\label{sec2}
\subsection{$l=0$ manifold}
\label{sec2a}
Let us consider the $l=0$ manifold. Its basis is formed by the set of ground states of each site, which we denote as $\{\ket{j}\}$, where the index $j=1,...,N$ labels the sites according to Fig.~ \ref{esquemaribbon}. Since the ground state wave functions have no angular dependence, the tunneling couplings are always real and only depend on the separation between sites, typically showing a fast decay as it is increased. Thus, it is a very good approximation to consider that only the central site of each cell is tunneled-coupled to its four nearest neighbours, with a tunneling rate $J$. Introducing the tunneling operators $\hat{a}^{\dagger}_k\hat{a}_j\ket{h}=\ket{k}\delta_{jh}$, where $k,j,h$ are labels denoting sites, the Hamiltonian for the $l=0$ manifold of a ribbon with $n$ cells can be written as
\begin{align}
\hat{H}_0&=-\hbar J \sum_{i=1}^{n}\left[(\hat{a}^{\dagger}_{3i-2}\hat{a}_{3i}+
\hat{a}^{\dagger}_{3i-1}\hat{a}_{3i}+\right.\nonumber
\\
&\left.\hat{a}^{\dagger}_{3i+1}\hat{a}_{3i}
+\hat{a}^{\dagger}_{3i+2}\hat{a}_{3i})+h.c.\right].
\label{BHl0}
\end{align}
To derive expressions for the ELS of the ribbon, we first consider a system of three sites $L,C,R$ (from left, central and right), with their ground states $\ket{L}$ and $\ket{R}$ equally coupled to $\ket{C}$ and decoupled between each other. This system has an eigenstate of 0 energy $\ket{D}=\frac{1}{\sqrt{2}}(\ket{L}+e^{i\pi}\ket{R})$, the SDS, which is decoupled from the state $\ket{C}$. In a ribbon of $n$ cells, ELS can be found by combining three-site SDS with population only on the edge sites. Thus, after setting them as initial states, the central sites $\{\ket{3i}\}$ ($i=1,...,n$) will remain unpopulated along the time evolution. In a single five-site cell there are 3 independent possibilities for such combinations, which correspond to the different ways in which the wave function at two of the outer sites can have a $\pi$ phase difference with respect to the other two. For $n$ cells there are $2^n+1$ possible ELS with equal population in the external sites,
\begin{align}
&\ket{D_k}_{l=0}=\frac{1}{\sqrt{2(n+1)}}(\ket{1}+e^{i\pi}\ket{2}+\nonumber\\
&\sum_{j=1}^n(-1)^{B_k^j(n)}(\ket{3j+1}+e^{i\pi}\ket{3j+2}));k=0,...,2^n-1\nonumber\\
&\ket{D_{2^n}}_{l=0}=\frac{1}{\sqrt{2(n+1)}}\sum_{j=0}^n e^{j\cdot i\pi}(\ket{3j+1}+\ket{3j+2}),
\label{darksl0}
\end{align}
where $B_k^j(n)$ is the $j$th digit (starting from the left) of the binary representation of $k$ using a total of $n$ digits. For instance, $B_7(n=4)=0111$ and $B_7^1(n=4)=0$. It can be checked that all the states \eqref{darksl0} are eigenstates of $\hat{H}_0$ with 0 energy.

\begin{figure}[t!]
\centering
\includegraphics[width=\linewidth]{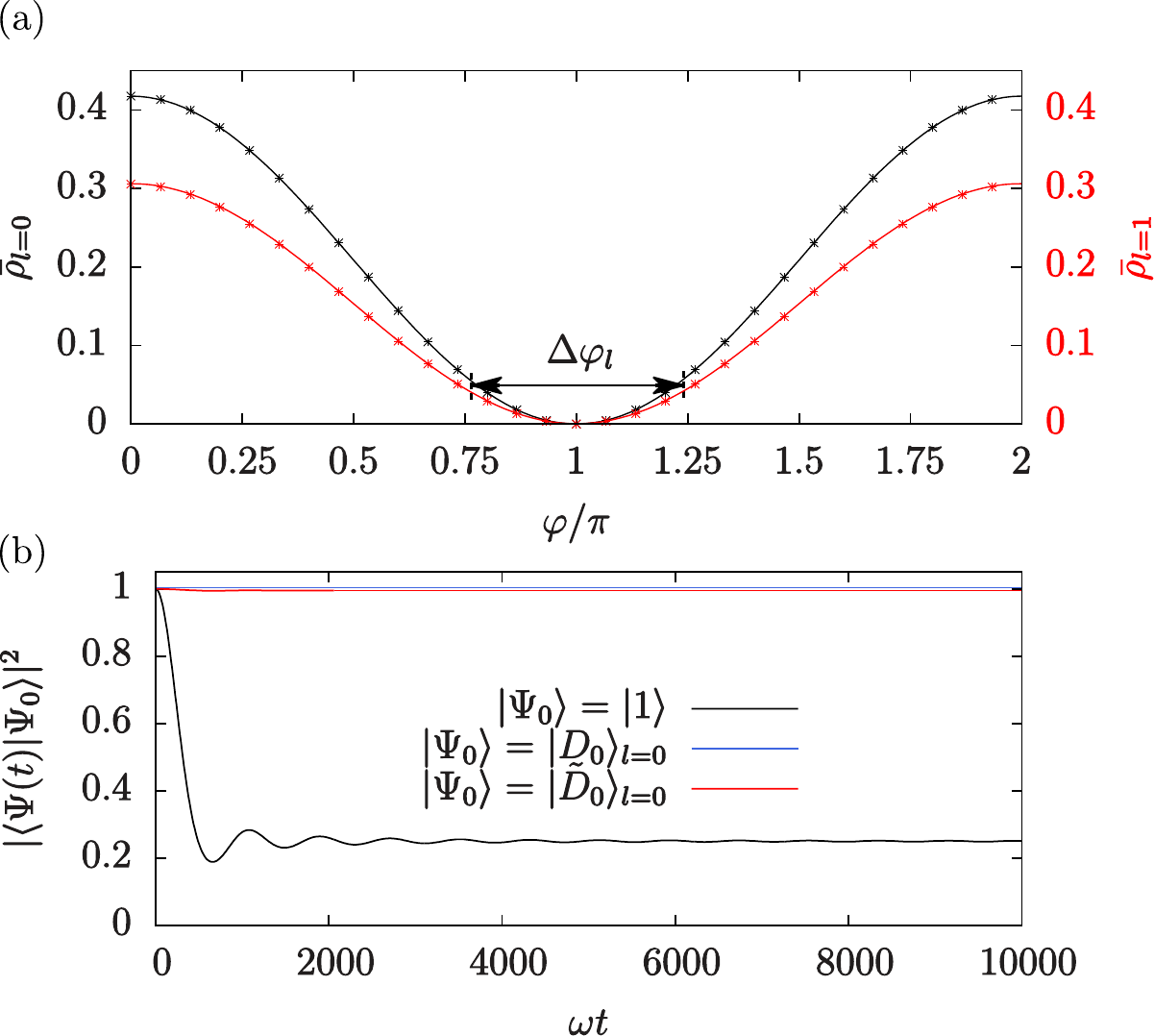}
\caption{(a) Average population of the central sites of a two-cell ribbon over a time $T=1000 \omega^{-1}$ when choosing as initial state $\ket{\Psi(\varphi)}_{l=0}$ (black) and $\ket{\Psi(\varphi)}_{l=1}$ (red). Solid lines and points correspond to the results obtained with the few-state models and full numerical integration of the 2D Schr\"odinger equation, respectively. (b) Time evolution of the population of different initial states $\ket{\Psi_0}$ under the action of $\hat{H}_0$ in a ribbon of 100 cells.}
\label{centralrho}
\end{figure}

To numerically check the existence of the ELS predicted with the few-state model of the $l=0$ manifold \eqref{BHl0}, we consider the following trial state in a ribbon of 2 cells
\begin{equation}
\ket{\Psi(\varphi)}_{l=0}=\frac{1}{\sqrt{6}}\sum_{j=0}^2 (\ket{3j+1}+e^{i\varphi}\ket{3j+2})
\label{triall0}
\end{equation} 
as initial state and find its time evolution both with the few-state model and by direct integration of the 2D Schr\"odinger equation. We then compute the average value of the populations of the central sites over the total time $T$, $\bar{\rho}_{l=0}(\varphi)=\sum_{i=1,2}\frac{1}{T}\int_0^Tdt\left|\braket{3i|e^{-i\hat{H}t/\hbar}|{\Psi}(\varphi)}_{l=0}\right|^2$. We have performed the calculations taking a distance between tunneled-coupled sites $d=5\sigma$, with $\sigma=\sqrt{\hbar/m\omega}$ ($m$ is the mass of the atom). The results are shown in Fig.~\ref{centralrho} (a). As predicted by the model, the average population of the central sites has a broad resonance at $\varphi=\pi$, since then the trial state is $\ket{D_0}_{l=0}$. For a range of phase differences $\Delta\varphi_{l=0}=0.4\pi$ around the minimum, the average population of the central sites remains below $0.05$. Thus, the ELS is robust against variations of the relative phases in the initial state. We also find that there is an excellent agreement between the model and the numerical integration.
\\
We have also checked the robustness of the ELS against local perturbations. In order to do so, we consider a ribbon of $n=100$ cells and set as initial states $\ket{1}$, $\ket{D_0}_{l=0}$ and this latter state with a defect, i.e. site 1 empty, which we denote as $\ket{\tilde{D}_0}_{l=0}$. In Fig.~\ref{centralrho} (b) we plot the time evolution of the populations of the initial states. While the single site state decays, the ELS with a defect maintains its population almost equal to 1 all along the time evolution, just like the original ELS. Thus, we conclude that the ELS is robust against local perturbations.
 
\begin{figure}[t!]
\centering
\includegraphics[width=\linewidth]{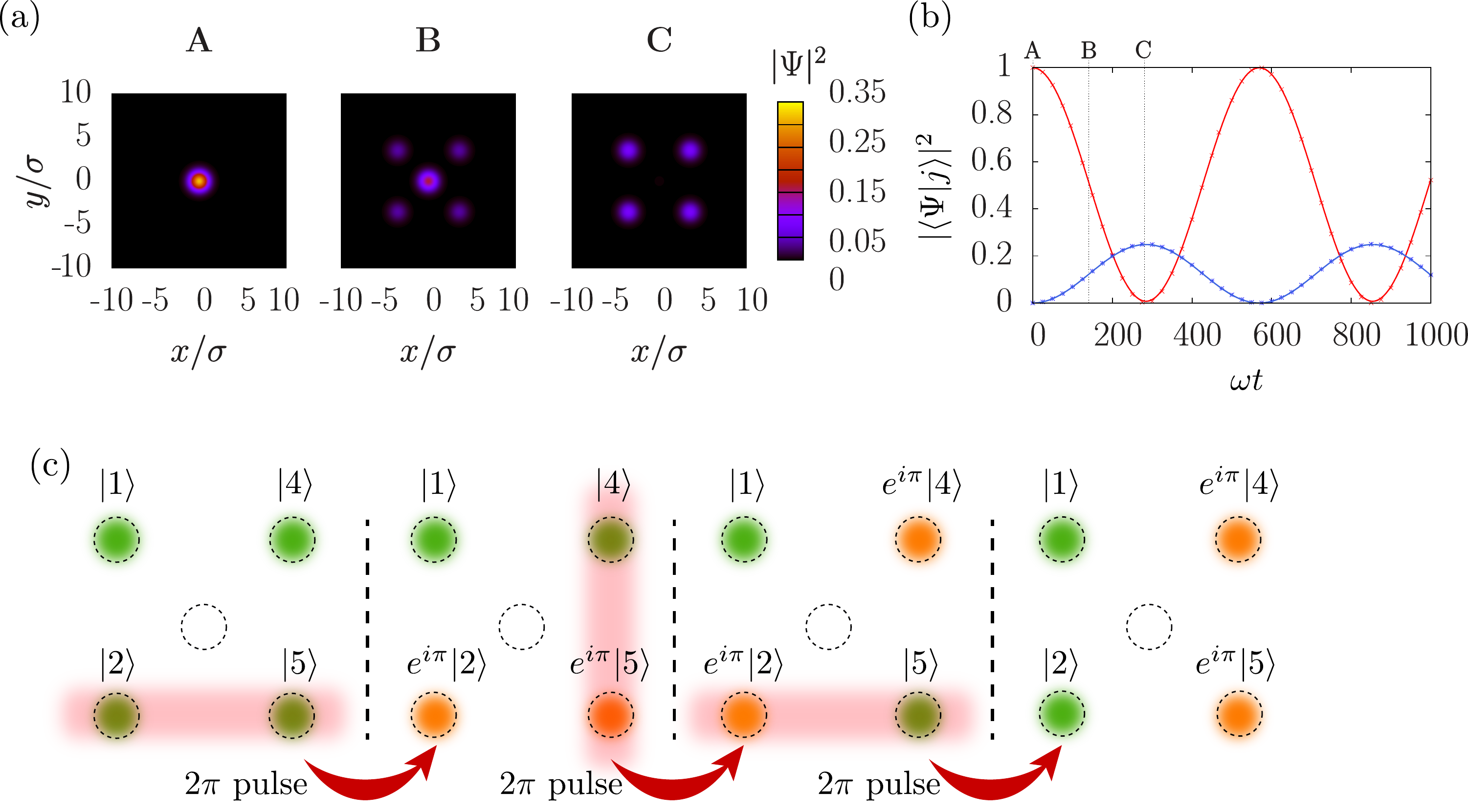}
\caption{(a) Density plots of the atomic wave function in a single cell of the considered ribbon when: the atom is located on the central site (A), some population has been transferred to the outer sites (B) and the central site is unpopulated (C). The separation between tunneled-coupled sites is $d=5\sigma$. (b) Time evolution of the populations of the central (red) and outer (blue) sites of the cell during the Rabi-type oscillations. Continuous lines correspond to the full integration of the Schr\"odinger equation and dots to the results obtained with the few-state Hamiltonian \eqref{BHl0}. (c) By illuminating two sites with a pulse of area $2\pi$, it is possible to induce $\pi$ phase changes on their local functions. In the leftmost sketch this is done at the instant C, giving rise to an ELS. The other panels show how sequential application of pulses allows to switch between the different ELS.}
\label{loadingswitching}
\end{figure}

The implementation of the ELS \eqref{darksl0}, restricted to a single cell for simplicity, is illustrated and supported by numerical simulations in Fig.~ \ref{loadingswitching} (a),(b). Initially, only the central site is populated (A). When left to evolve, the atom undergoes Rabi-type oscillations in which the population is transferred equally and with the same phase to the four outer sites (B). The populations of the different sites during this process are shown in Fig.~ \ref{loadingswitching} (b). When all the population has been transferred to the outer sites (C), one can apply a laser pulse of area $2\pi$ that induces a $\pi$ change of phase in the states localized in all the sites of either the upper or the lower row, as shown in the leftmost panel of Fig.~ \ref{loadingswitching} (c). In this way, the system will be transferred to an ELS and remain there. Once the system has been prepared in this initial ELS, by applying localized $2\pi$ pulses in one column or row it is possible to induce $\pi$ phase changes in the states localized in any desired pair of sites of the ribbon, and thus switch from one ELS to another, as shown in Fig.~ \ref{loadingswitching} (c). Thus, these phase-change processes allow to explore the whole $l=0$ ELS subspace. Also, by applying $\pi$ pulses to pairs of sites, one could create equally weighted superpositions of ELS.  
This kind of ELS can also be implemented in more complex geometries, as shown in Fig.~ \ref{finalplot} (a). By populating equally all the edge sites but the four corners and setting $\pi$ phase differences between local states of adjacent edge sites, quantum interference would prevent population transfer to any tunneled-coupled site and an ELS would be created.  
\subsection{$l=1$ manifold}
\label{sec2b}
\begin{figure}[t!]
\centering
\includegraphics[width=0.8\linewidth]{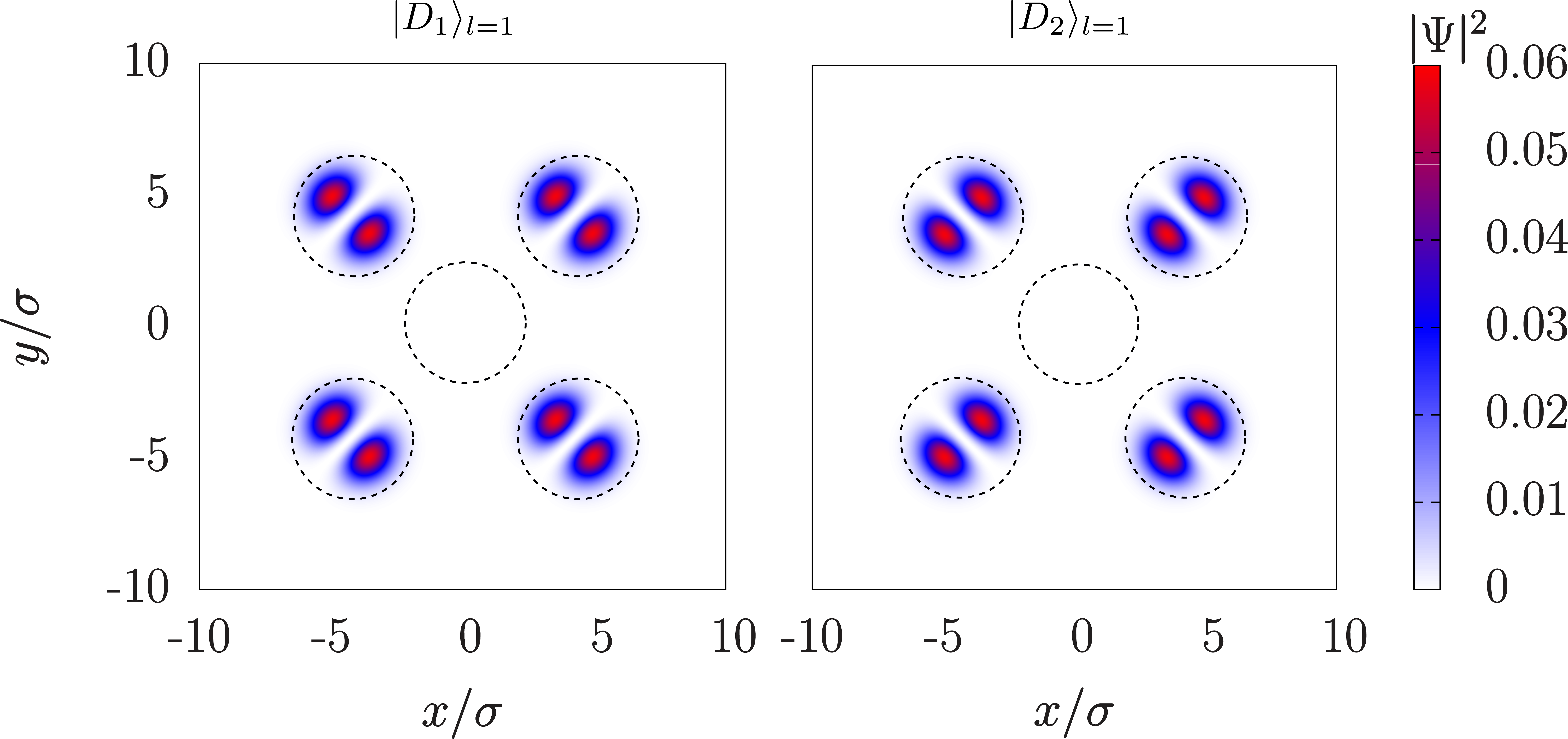}
\caption{Density profile of the two ELS \eqref{darksl1} of the $l=1$ manifold restricted to one cell of a ribbon with $d=6\sigma$. For each state, the wave function has a nodal line with the same orientation at all sites, giving rise to global chirality.}
\label{edgel1}
\end{figure}

We now consider the $l=1$ manifold of the 2D ribbon shown in Fig.~ \ref{esquemaribbon}. The basis of this manifold is formed by the set of first excited eigenstates of each site with winding number $m=\pm 1$, denoted as $\{\ket{j,+},\ket{j,-}\}$, which could be implemented according to \citep{l4}. Their wave functions are of the form $\braket{\vec{r}|j,\pm}\sim \Psi(r_j)e^{\pm i(\phi_j-\phi_0)}$, where $(r_j,\phi_j)$ are the polar coordinates with origin at the position of the $j$th site and $\phi_0$ is a free phase parameter. Because of the angular dependence of the wave functions, the couplings between the different states depend both on the distance between sites and on their relative orientation, and can in general take complex values. Using symmetry considerations, it can be shown \cite{geometricallyinduced} that for a given separation between two sites $j,k$ there are only three independent coupling rates: a self-coupling rate between the two eigenstates of a site, $J_1\sim|\braket{j,\pm|\hat{H}|j,\mp}|$, and two cross-coupling rates between the eigenstates of different sites with equal or different winding number, $J_2\sim|\braket{j,\pm|\hat{H}|k,\pm}|$ and $J_3\sim|\braket{j,\pm|\hat{H}|k,\mp}|$. The phases of the complex tunneling amplitudes depend on $\phi_0$ and the winding numbers of the states involved in the coupling. As for $l=0$, the coupling rates decay fast with the inter-site separation, so again we will consider that only the central sites are tunneled-coupled to their neighbours. By setting $\phi_0=-\pi/4$, all the couplings between the sites $3i-2\leftrightarrow3i$ and $3i\leftrightarrow 3i+2$ are real, whereas between the sites $3i-1\leftrightarrow3i$ and $3i\leftrightarrow 3i+1$ the couplings involving winding number change acquire $\pi$ phases \cite{geometricallyinduced}. By defining the tunneling operators $\hat{a}^{\dagger}_{j,\alpha''}\hat{a}_{k,\alpha'}\ket{h,\alpha}=\ket{j,\alpha''}\delta_{kh}\delta_{\alpha'\alpha
}$, where $\alpha=\pm$ is the winding number, the Hamiltonian of a ribbon with $n$ cells reads
\begin{align}
&\hat{H}_1=\nonumber
\\
&-\hbar\sum_{i=1}^n\sum_{\alpha,\alpha'=\pm 1}(U_1)_{\alpha\alpha'}(\hat{a}^{\dagger}_{3i,\alpha}\hat{a}_{3i-1,\alpha'}+\hat{a}^{\dagger}_{3i,\alpha}\hat{a}_{3i+1,\alpha'})\nonumber
\\
&-\hbar\sum_{i=1}^n\sum_{\alpha,\alpha'=\pm 1}(U_2)_{\alpha\alpha'}(\hat{a}^{\dagger}_{3i,\alpha}\hat{a}_{3i-2,\alpha'}+\hat{a}^{\dagger}_{3i,\alpha}\hat{a}_{3i+2,\alpha'})\nonumber
\\
&-\hbar\sum_{\alpha,\alpha'}(S_1)_{\alpha\alpha'}(\hat{a}^{\dagger}_{1,\alpha}\hat{a}_{1,\alpha'}+\hat{a}^{\dagger}_{N-1,\alpha}\hat{a}_{N-1,\alpha'})\nonumber
\\
&-\hbar\sum_{\alpha,\alpha'}(S_2)_{\alpha\alpha'}(\hat{a}^{\dagger}_{2,\alpha}\hat{a}_{2,\alpha'}+\hat{a}^{\dagger}_{N,\alpha}\hat{a}_{N,\alpha'})+h.c.,
\label{BHl1}
\end{align}
with coupling matrices
\begin{subequations}
\begin{eqnarray}
U_1=\begin{pmatrix}
J_2&J_3e^{-i\pi}\\
J_3e^{i\pi}&J_2
\end{pmatrix}
;\quad
U_2=\begin{pmatrix}
J_2&J_3\\
J_3&J_2
\end{pmatrix}
\\
S_1=\begin{pmatrix}
0&J_1\\
J_1&0
\end{pmatrix}
;\quad
S_2=\begin{pmatrix}
0&J_1e^{-i\pi}\\
J_1e^{i\pi}&0
\end{pmatrix}
.
\end{eqnarray}
\end{subequations}
Note that the complex number sum rule for the contributions to the self-coupling makes this term vanish at all the sites but the four corners of the ribbon \cite{geometricallyinduced}. In addition, $|J_2|\approx |J_3|$ and $|J_1|\ll|J_{2}|,|J_{3}|$, so in a first approximation the self-couplings can be neglected \cite{geometricallyinduced}. Within this approximation, a system of three in-line sites $L$, $C$ and $R$ ($L$ and $R$ equally separated from $C$) with $l=1$ local eigenstates $\ket{L,\pm}$, $\ket{C,\pm}$ and $\ket{R,\pm}$ has two SDS of 0 energy, $\ket{D+}=\frac{1}{2}(\ket{L,+}+\ket{L,-}-\ket{R,+}-\ket{R,-})$ and $\ket{D-}=\frac{1}{2}(\ket{L,+}-\ket{L,-}-\ket{R,+}+\ket{R,-})$. On a ribbon, ELS can be implemented by setting the SDS $\ket{D\pm}$ along the lines $3i-2\leftrightarrow 3i\leftrightarrow 3i+2$ and $3i-1\leftrightarrow 3i\leftrightarrow 3i+1$. Among the many possibilities to do so, two are particularly interesting because the orientation of the nodal lines of the wavefunction gives rise to global chirality, as shown in Fig.~ \ref{edgel1}. These two ELS read
\begin{subequations}
\begin{align}
\ket{D_1}_{l=1}&=\frac{1}{\sqrt{4(n+1)}}\sum_{j=0}^n \big[(\ket{3j+1,+}+\ket{3j+1,-})\nonumber\\
&+e^{i\pi}(\ket{3j+2,+}+\ket{3j+2,-})\big]\\
\ket{D_2}_{l=1}&=\frac{1}{\sqrt{4(n+1)}}\sum_{j=0}^n \big[(\ket{3j+1,+}-\ket{3j+1,-})+\nonumber\\
&+e^{i\pi}(\ket{3j+2,+}-\ket{3j+2,-})\big].
\end{align}
\label{darksl1}
\end{subequations}
To test the ELS \eqref{darksl1}, we consider the following trial state in a ribbon of 2 cells
\begin{align}
\ket{\Psi(\varphi)}_{l=1}&=\frac{1}{\sqrt{12}}\sum_{j=0}^2 \big[(\ket{3j+1,+}\ket{3j+1,-})\nonumber\\
&+e^{i\varphi}(\ket{3j+2,+}+\ket{3j+2,-})\big] 
\end{align}
\begin{figure}[t!]
\centering
\includegraphics[width=\linewidth]{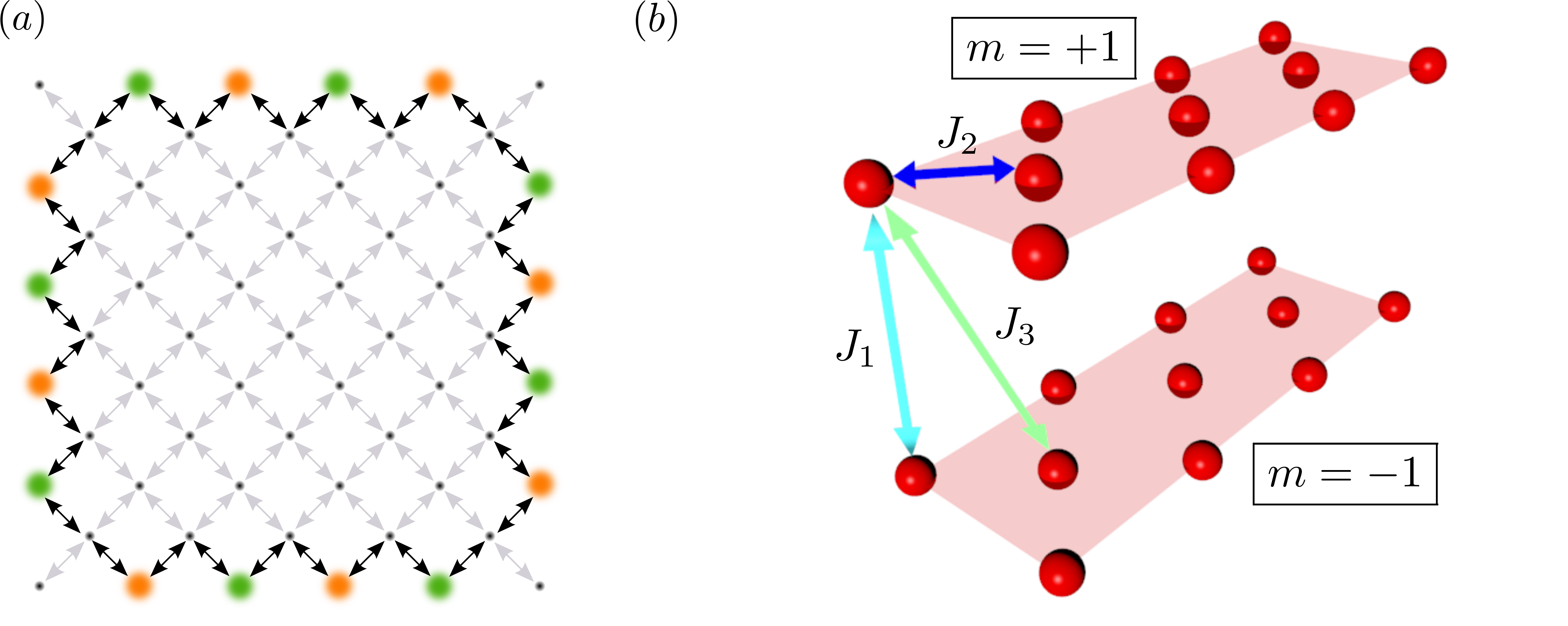}
\caption{(a) ELS constructed from SDS of $l=0$ on a tilted square lattice, with all edge sites but the corners equally populated. As indicated by the alternating orange and green colourings, the atomic states at adjacent sites have $\pi$ relative phases. Double arrows indicate nearest neighbour tunneling couplings, with the ones involved in the SDS highlighted in black. (b) Schematics of the extension of the optical ribbon into the synthetic dimension given by the winding number in the $l=1$ case. Each red dot can host only one state, and tunneling-coupling (represented by the double arrows) can occur through sites in the same or different winding number planes.}
\label{finalplot}
\end{figure}
\hspace{-0.2 cm}and compute the average population of the central-site states $\bar{\rho}_{l=1}(\varphi)$, see Fig.~ \ref{centralrho} (a). As expected, $\bar{\rho}_{l=1}(\varphi)$ has a resonance at $\varphi=\pi$, where it becomes 0, whose width is even larger than in the $l=0$ case: the average population of the central sites remains under $0.05$ for $\Delta\varphi_{l=1}=0.5\pi$. Thus, the ELS constructed from combinations of $l=1$ SDS are also robust. We have taken a longer separation between sites $d=6\sigma$ because the wave functions of the first excited states are more extended than the ground state ones. Note that the simulations account for the self-coupling terms. Thus, although the states \eqref{darksl1} are eigenstates of $\hat{H}_1$ only for $J_1=0$, they work well as ELS in the sense that they do not populate the central sites for $J_1\neq 0$. The fact that they are not actual eigenstates is manifested in small oscillations of the populations of the occupied edge sites. We have checked that in the longest time scale of the system, typically $\frac{2\pi}{J_1}\sim 10^4\omega^{-1}$, these oscillations decay as the number of cells in the ribbon increases. Hence, they may be considered as finite-size effects due to the corners of the ribbon, with no relevance in the limit of large number of cells. Following the same procedure as in the case of the $l=0$ manifold, we have checked that the $l=1$ ELS are also robust against local perturbations. 

So far, we have considered the winding number associated to the OAM as a degree of freedom that increases the number of states per site. However, one could also conceive it as a synthetic dimension, see Fig.~ \ref{finalplot} (b). In this picture, the 2D ribbon splits into two layers, each containing only states with a well-defined winding number. 
Within each layer, the central sites are connected to their nearest neighbours through $J_2$ cross-couplings.
The two layers are connected by cross-couplings $\pm J_3$ between central and edge sites, and at the four corners there are vertical connections between edge sites corresponding to self-couplings $\pm J_1$. This approach could open the possibility of using 2D optical lattices to simulate three-dimensional systems, in which the high variety of possible closed paths could yield non-abelian artificial gauge fields \cite{goldman_etal_review}.
\section{Conclusions}
\label{sec3}
In this work we have shown that quantum interference effects can be used to generate robust single atom ELS in arbitrarily large optical ribbons. These ELS may be built within the manifolds of local eigenstates with total OAM $l=0$ or $l=1$. For $l=0$, the different ELS could be easily coupled with laser pulses, allowing to induce oscillations between global eigenstates of the ribbon. Also, the versatility of the three-site SDS in which the ELS are based could be used to implement similar states in other geometries. For $l=1$, conceiving the winding number as an extra synthetic dimension could lead to quantum-simulate exotic three-dimensional lattices or synthetic gauge fields. A possible extension of this work is the study of the role of the nonlinearity in similar configurations.
\acknowledgements
The authors gratefully acknowledge financial support through the Spanish Ministry of Economy and 
Competitiveness (MINECO) (Contract No. FIS2014-57460-P, FIS2013-46768-P, Severo Ochoa
Excellence Grant SEV-2015-0522) and the Catalan Government (Contract No. SGR2014-1639 and SGR 2014-8742). G.P and J.P. acknowledge financial support from the MINECO through the Grants No. BES-2015-073772 and BES-2012-053447, respectively. M. L. acknowledges ERC AdG OSYRIS, EU FETPRO QUIC, and EU IP SIQS.


\begin{thebibliography}{99}
\bibitem{llibreveronica} M. Lewenstein, A. Sanpera, and V. Ahufinger, \textit{Ultracold Atoms in
Optical Lattices: Simulating Quantum Many-Body Systems} (Oxford Univ. Press, Oxford, 2012).

\bibitem{reviewbloch} I. Bloch, J. Dalibard, and S. Nascimb\`ene, \textit{Nat. Phys.} \textbf{8}, 267-276
(2012).

\bibitem{jaksch_zoller} D. Jaksch and P. Zoller, \textit{New J. Phys.} \textbf{5}, 56 (2003).

\bibitem{osterloh2005} K. Osterloh, M. Baig, L. Santos, P. Zoller, and M. Lewenstein, \textit{Phys. Rev. Lett.} \textbf{95}, 010403 (2005).

\bibitem{mueller} E. J. Mueller, \textit{Phys. Rev. A} \textbf{70}, 041603(R) (2004).

\bibitem{gerbier_dalibard} F. Gerbier and J. Dalibard, \textit{New J. Phys.} \textbf{12}, 033007 (2010).

\bibitem{eckert2010} A. Eckardt, P. Hauke, P. Soltan-Panahi, C. Becker, K. Sengstock, and M. Lewenstein, \textit{Europhys. Lett.} \textbf{89}, 10010 (2010).

\bibitem{struck2012} J. Struck, C. \"Olschl\"ager, M. Weinberg, P. Hauke, J. Simonet, A. Eckardt, M. Lewenstein, K. Sengstock, and P. Windpassinger, \textit{Phys. Rev. Lett.} \textbf{108}, 225304 (2012). 

\bibitem{hauke2012} P. Hauke, O. Tieleman, A. Celi, C. \"Olschl\"ager, J. Simonet, J. Struck, M. Weinberg, P. Windpassinger, K. Sengstock, M. Lewenstein, and A. Eckardt, \textit{Phys. Rev. Lett.} \textbf{109}, 145301 (2012).

\bibitem{gediminasRMP} J. Dalibard, F. Gerbier, G. Juzeli\={u}nas, and P. \"Ohberg, \textit{Rev. Mod. Phys.} \textbf{83}, 1523 (2011).

\bibitem{goldman_etal_review} N. Goldman, G. Juzeli\={u}nas, P. \"Ohberg, and I. B. Spielman, \textit{Rep.
Prog. Phys.} \textbf{77}, 126401 (2014).

\bibitem{shakingreview} N. Goldman and J. Dalibard, \textit{Phys. Rev. X} \textbf{4}, 031027 (2014).

\bibitem{chernumber}  M. Aidelsburger, M. Atala, M. Lohse, J.T. Barreiro, B. Paredes, and I. Bloch, \textit{Nat. Phys.} \textbf{11}, 162-166 (2015).

\bibitem{BECstrongfield}  C. J. Kennedy,	W. C. Burton, W. C. Chung, and W. Ketterle \textit{Nat. Phys.} \textbf{11}, 859-864 (2015).

\bibitem{currentbosons} M. Atala, M. Aidelsburger, M. Lohse, J.T. Barreiro, B. Paredes, and I. Bloch, \textit{Nat. Phys.} \textbf{10}, 588-593 (2014).

\bibitem{hofstader} M. Aidelsburger, M. Atala, M. Lohse, J.T. Barreiro, B. Paredes, and I. Bloch, \textit{Phys. Rev. Lett.}, \textbf{111}, 185301 (2013).

\bibitem{harper} H. Miyake, G.A. Siviloglou, C.J. Kennedy, W.C. Burton, and W. Ketterle, \textit{Phys. Rev. Lett.}, \textbf{111}, 185302 (2013).

\bibitem{edgebosons} B. K. Stuhl, H.I. Lu, L.M. Aycock, D. Genkina, and I.B. Spielman, \textit{Science} \textbf{349}, 1514-1518 (2015).

\bibitem{edgefermions} M. Mancini, G. Pagano, G. Cappellini, L. Livi, M. Rider, J. Catani, C. Sias, 
P. Zoller, M. Inguscio, M. Dalmonte, and L. Fallani, \textit{Science} \textbf{349}, 1510-1513 (2015). 

\bibitem{syntheticdimensions} A. Celi, P. Massignan, J. Ruseckas, N. Goldman, I.B. Spielman, G. Juzeli\={u}nas, and M. Lewenstein, \textit{Phys. Rev. Lett.}, \textbf{112}, 043001 (2014). 

\bibitem{edgequantuminfo} N.Y. Yao, C.R. Laumann, A.V. Gorshkov, H. Weimer, L. Jiang, J.I. Cirac,
P. Zoller, and M.D. Lukin, \textit{Nat. Commun.} \textbf{4}, 1585 (2013).

\bibitem{edgequantuminfo2} C. Dlaska, B. Vermersch, and P. Zoller, \textit{Quantum Sci. Tech.} \textbf{2}, 015001 (2017).
 
\bibitem{spatialdark} K. Eckert, M. Lewenstein, R. Corbal\'an, G. Birkl, W. Ertmer, and J. Mompart, \textit{Phys. Rev. A} \textbf{70}, 023606 (2004).

\bibitem{reviewsap} R. Menchon--Enrich, A. Benseny, V. Ahufinger, A.D. Greentree, Th. Busch, and J Mompart, \textit{Rep. Prog. Phys.} \textbf{79}, 074401 (2016).


\bibitem{painting1} S. K. Schnelle, E. D. van Ooijen, M. J. Davis, N. R. Heckenberg, and H. Rubinsztein-Dunlop, \textit{Opt. Express} \textbf{16}, 1405 (2008).

\bibitem{painting2}  N. Houston, E. Riis and A.S. Arnold, \textit{J. Phys. B} \textbf{41}, 211001 (2008).

\bibitem{painting3} K. Henderson, C. Ryu, C. MacCormick and M.G. Boshier, \textit{New J. Phys.} \textbf{11}, 043030 (2009).

\bibitem{painting4} A. S. Arnold, \textit{Opt. Lett.} \textbf{37}, 2505 (2012).

\bibitem{slm1} F. Nogrette, H. Labuhn, S. Ravets, D. Barredo, L. B\'eguin, A. Vernier, T. Lahaye, and A. Browaeys, \textit{Phys. Rev. X}, \textbf{4}, 021034 (2014).

\bibitem{slm2} H. Tamura, T. Unakami, J. He, Y. Miyamoto, and K. Nakagawa, \textit{Opt. Express}, \textbf{24}, 8132-8141 (2016).

\bibitem{geometricallyinduced} J. Polo, J. Mompart, and V. Ahufinger, \textit{Phys. Rev. A} \textbf{93}, 033613 (2016).

\bibitem{l1} S. L. Zhang and Q. Zhou, \textit{Phys. Rev. A} \textbf{90}, 051601(R) (2014).

\bibitem{l2} C. Str\"ater and A. Eckardt, \textit{Phys. Rev. A} \textbf{91}, 053602 (2015).

\bibitem{l3} S. L. Zhang, L. J. Lang, and Q. Zhou, \textit{Phys. Rev. Lett.} \textbf{115}, 225301 (2015).

\bibitem{l4} A. Kiely, A. Benseny, T. Busch and A. Ruschhaupt, \textit{J. Phys. B} \textbf{49}, 215003 (2016).

\bibitem{l5} T. Sowi\'nski, \textit{Phys. Rev. Lett.} \textbf{108}, 165301, (2012).

\bibitem{l6} G. Wirth, M. \"Olschl\"ager, and A. Hemmerich, \textit{Nat. Phys.} \textbf{7}, 147 (2011).

\bibitem{l7} T. Kock, C. Hippler, A. Ewerbeck, and A. Hemmerich, \textit{J. Phys. B: At. Mol. Opt. Phys.} \textbf{49}, 042001 (2016).

\bibitem{l8} Z. Xu, L. You, A. Hemmerich, and W. V. Liu, \textit{Phys. Rev. Lett.} \text{117}, 085301 (2016).

\bibitem{l9} C. V. Parker, L. C. Ha, and C. Chin, \textit{Nat. Phys.} \textbf{9}, 769 (2013).

\bibitem{l10} M. A. Khamehchi, C.  Qu, M. E. Mossman, C. Zhang, and P. Engels, \textit{Nat. Commun.} \textbf{7}, 10867 (2016).

\bibitem{l11} T. M\"uller, S. F\"olling, A. Widera, and I. Bloch, \textit{Phys. Rev. Lett.} \textbf{99}, 200405 (2007).

\bibitem{nontrivialtopology} O. Boada, A. Celi, J. Rodr\'iguez-Laguna, J.I. Latorre, and M. Lewenstein, \textit{New J. Phys.} \textbf{17}, 045007 (2015).

\bibitem{synthdimphotonic} T. Ozawa and I. Carusotto, \textit{Phys. Rev. Lett.} \textbf{118}, 013601 (2017).

\end{thebibliography}
\end{document}